**Flowering of Developable 2D Crystal Shapes in Closed, Fluid Membranes**

Hao Wan,[1] Geunwoong Jeon,[2] Weiyue Xin,[3] Gregory M. Grason,[1] and Maria M. Santore[1,*]

1. Department of Polymer Science and Engineering, University of Massachusetts, 120 Governors Drive, Amherst, MA 01003
2. Department of Physics, University of Massachusetts, 710 N. Pleasant Street, Amherst, MA 01003
3. Department of Chemical Engineering, University of Massachusetts, 686 N. Pleasant Street, Amherst, MA 01003

The morphologies of two-dimensional (2D) crystals, nucleated, grown, and integrated within 2D elastic fluids, for instance in giant vesicle membranes, are dictated by an interplay of mechanics, permeability, and thermal contraction. Mitigation of solid strain drives formation of crystals with developable shapes (e.g. planar or cylindrical) that expel Gaussian curvature into the 2D fluid. However, upon cooling to grow the crystals, large vesicles sustain greater inflation and tension because their small area to volume ratio slows water permeation.  As a result, more elaborate shapes, for instance flowers with bendable but inextensible petals form on large vesicles despite their more gradual curvature, while small vesicles harbor compact planar crystals. This size dependence runs counter to the known cumulative growth of strain energy of 2D colloidal crystals on rigid spherical templates. This interplay of intra-membrane mechanics and processing points to the scalable production of flexible molecular crystals of controllable complex shape.

**MAIN**

The importance of ultrathin and flexible materials motivates a focus on nanometrically thin bendable crystals whose sheet-like character and large lateral areas qualify them as 2D solids. Experiments, simulations and theory show that a 2D crystal growing on a fixed spherical template avoids topological defects via a stress-triggered boundary instability,[3, 4] producing a progression from compact, hexagonal domains to highly anisotropic protrusions and stripes which reduce the energetic cost of Gaussian curvature.[5, 6, 7, 8] The characteristic domain length scale is controlled by the interplay between in-plane stresses, line energy, and sphere radius.[5, 6, 9] Thus on smaller templates with greater curvature, the transition from compact to striped/protruding morphology occurs earlier during crystal growth, with narrower protrusions and stripes emanating from smaller compact domain cores.

Here we show how the morphologies of 2D crystals growing within an elastic 2D fluid having a flexible shape and a closed topology, for instance solid domains integrated into the otherwise fluid membrane of a giant unilamellar vesicle, such as that shown schematically in Figure 1A, are controlled by a fundamentally distinct mechanism. The in-plane solid elasticity of 2D crystals, i.e. a preference for flatness or cylindrical bending,[10, 11] expels Gaussian curvature to the 2D fluid,[12] leading to a complex interplay between the shape and morphology of the crystal and the system's global shape and bending energy. We employ fluid $L_\alpha$ phase phospholipid bilayer membranes[13, 14] containing integrated solid membrane domains as a platform to explore how 2D crystals, growing within a curved 2D elastic fluid, adjust both their morphology and curvature to minimize the total energy. The size sensitivity of the resulting morphological instability runs counter to that for crystallization on rigid spherical templates.[5, 6, 9] Further, thermal membrane contractions and water permeation from the vesicle tune membrane tension and energy to scalably produce a vesicle size dependence of crystal morphologies. When thermal history and osmotic preconditioning are fixed for the entire suspension, more elaborate crystal shapes grow systematically on larger more gradually curved vesicles, at the same membrane composition.

**Emergent Shapes of Crystallized Domains**

In giant unilamellar vesicle membranes containing two or more phospholipids, ordered crystalline membrane domains, some with intricate shapes,[15] can coexist within the fluid membrane phase, called an $L_\alpha$ phase.[16, 17, 18, 19] In this work the $L_\alpha$ fluid membrane contains a mixture of l,2-dipalmitoyl-*sn*-glycero-3-phosphocholine (DPPC) and l,2-dioleoyl-*sn*-glycero-3-phosphocholine (DOPC), plus a fluorescent tracer lipid (l,2- dioleoyl-*sn*-glycero-3-phosphoethanolamine-*N*-(lissamine rhodamine B sulfonyl) (ammonium salt), Rh-DOPE) that is excluded from the nearly pure ordered DPPC solid domains, enabling their visualization in membranes containing both fluid and solid domains. The $L_\alpha$ fluid membrane can bend and stretch elastically and, because it can also shear freely,[20, 21, 22, 23] it assumes Gaussian curvature at a low cost. Unlike the fluid membrane, solid phospholipid domains possess identical crystalline order in both leaflets[24, 25, 26, 27, 28] and correspondingly non-zero 2D shear moduli,[20] imparting shear rigidity that limits access to shapes like spheres that have non-zero Gaussian curvature. Solid shear elasticity underlies the geometrically nonlinear coupling between Gaussian curvature and in-plane strain of 2D sheets.[10] While changes in mean curvature (i.e. bending) incur modest cost due to the nanometrically thin dimensions,[20, 23, 29, 30, 31, 32] non-zero Gaussian curvature in the solid leads to prohibitively large shear strains that grow with the much greater lateral dimensions of the 2D crystal.[9, 33] Notably, the formation of topological defects that form in some curved 2D crystals[34, 35, 36] can only relax some, but not all, of the large thermodynamic costs of Gaussian curvature.[37]

Starting with an overall membrane composition of 30/70 DPPC/DOPC molar ratio in the one phase region of the phase diagram in Figure 1B, cooling at an appropriate rate produces solid domains, one per vesicle, that first appear just below 32°C and grow progressively until at room temperature, they occupy area fraction $\phi = 0.12 - 0.14$ of the membrane, according to a mass balance on the room temperature tie line (Figure 1B). An example calculation can be found in the Supporting Information. A variety of solid shapes, all with 6-fold symmetry from convex hexagons and highly non-convex stars to flowers, are found with examples shown in Figure 1C. The room temperature solid area fractions measured from image analysis of 15-20 vesicles of each crystal shape type, summarized within Figure 1C and detailed in the SI, are independent of domain shape and agree with the mass balance on the room temperature tie line. Thus despite the varied morphologies of the solid domains, all the vesicle membranes adhere to the thermodynamic phase diagram, suggesting mechanisms other than nonequilibrium thermodynamics to explain the varied domain morphology.

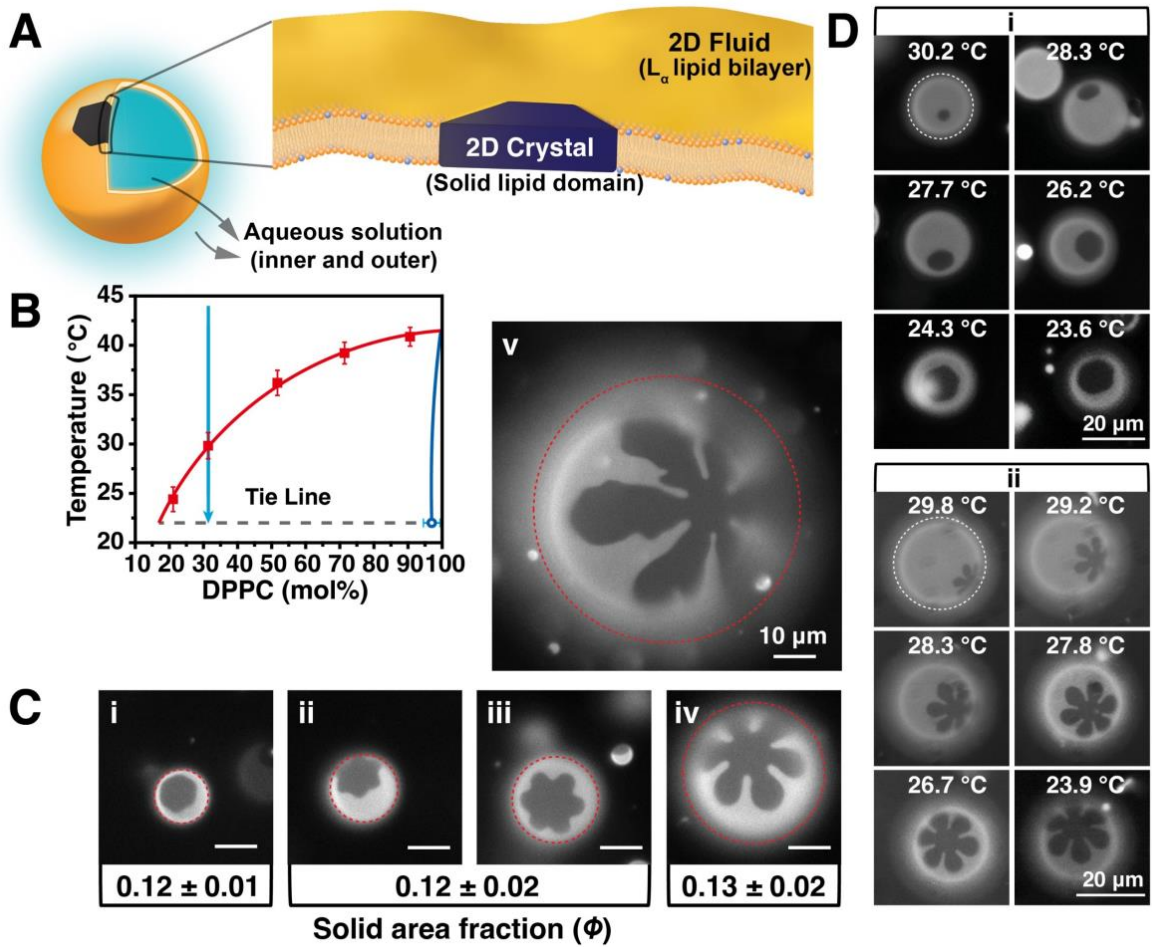

**Figure 1**. **A**. Schematic of solid crystal in a 2-D fluid comprising an $L_\alpha$ lipid bilayer.
**B**. Phase diagram and cooling trajectory for giant unilamellar vesicles membranes of lipid mixtures of DPPC and DOPC and < 0.1 mol % tracer. $L_\alpha$ phase boundary based on the first appearance of solid domains upon cooling at each composition. Solid datum approximates solid domains known to be nearly pure in DPPC.[1, 2]
**C.** Examples of DPPC solid crystal shapes seen on vesicles of different sizes, all in the same batch and shown with the same 10- μm scale bar. The solid crystalline domains are the dark shapes, while the $L_\alpha$ fluid is bright due to the fluorescent tracer. Solid area fractions for different shape types (convex hexagon, ninja star flower, simple flower) are included and represent averages of 10-15 vesicles of each shape type, detailed in SI. The red boundaries indicate the vesicle size, determined by a separate image taken a few moments apart focusing on the equatorial plane rather than the crystal-containing surface.
**D**. i and ii. Two series of vesicle images, each showing a progression for a single vesicle and its growing crystal shape during cooling. The first image in each series was acquired as soon as a crystal could be identified and focused. The rest of the images continue to room temperature. The two vesicles differ only in size. The dashed boundaries in each first image indicate the vesicle size, based on a separate image recorded at a slightly different time, focusing on the equatorial plane rather than the pattern.

Two unusual features are observed in these systems. First, as shown in two typical examples in Figure 1D, the initially discernable solid shape is preserved as domains grow. This observation suggests crystal growth and shape development by molecular addition at domain edges as opposed to aggregation of small domains to produce large ones. There is also a lack of classical dendritic growth instabilities since the shapes do not branch as they grow. Indeed the images of Figure 1D are consistent with single crystal growth. Second, evident in Figure 1C, there is a strong correlation between domain shape and vesicle size. More compact, convex domains are found on smaller vesicles and solid domains with more extended protrusions and elaborate flower shapes on the larger vesicles. This is particularly remarkable because all these vesicles have the same composition, same solid area fraction at room temperature, and were processed together to produce nucleation and growth in a single chamber with the same osmotic handling and thermal program. The particular vesicles in Figure 1C were selectively visualized by translating the microscope stage to focus on different vesicles.

**Vesicle Size Selects Crystal Shape**

The solid domain shapes comprise a continuum defined by the ratio of the circumradius to inradius, i.e. $\alpha = D_{outer}/D_{core}$. The nominal shape types, discernable by eye, in Figure 2A provide a convenient framework for classifying solid domains and always exhibit the $\alpha$ values in Figure 2A. For instance $\alpha$ varies from 1-1.15 for hexagons and convex domains that are less sharply faceted, up to >3.5 for serrated flowers. The categories of hexagons, ninja star flowers, simple flowers, and serrated flowers enabled us to establish, in Figure 2B the dependence of domain shape on vesicle size for over 330 vesicles in three separate runs, all with the same lipid composition, osmotic conditions and thermal history. Figure 2B establishes distinct vesicle size ranges where crystals of different shapes are found. The correlation is strong, with serrated flowers found only on the largest vesicles and for instance, compact hexagonal domains never seen on vesicles greater than 25 μm in diameter.

Remarkably, the most compact shapes are found on the small vesicles, i.e. those with the greatest curvature, while non-convex flowers are seen on larger vesicles, having smaller curvature. This correlation runs counter to predictions of Foppl von-Karman theory,[5, 9] phase-field modeling,[4, 7] and experiments of colloidal crystallization on fixed spherical templates[6] that show the threshold size for the transition from compact to anisotropic (e.g. stripe, branched or non-convex) domain shape *increases* with larger sphere radius.

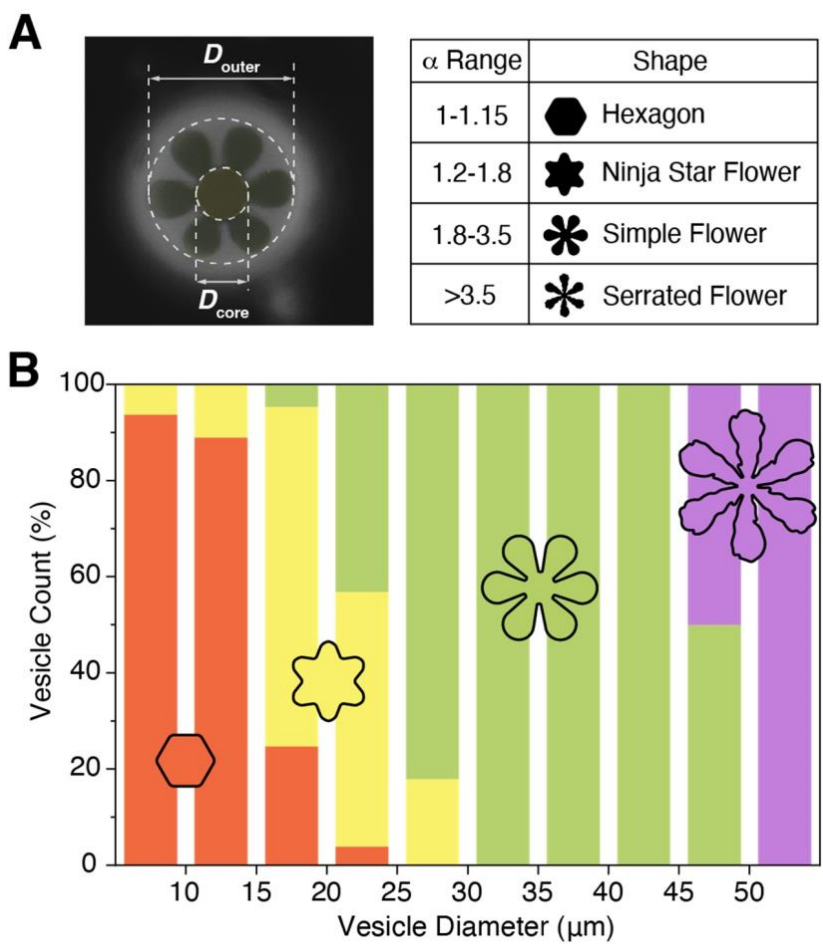

**Figure 2**. **A**. Schematic illustrating how $\alpha$ was measured and the range of $\alpha$ for different types of solid shapes. **B**. Summary of solid domain shapes found on 330 vesicles of different sizes. More than 100 vesicles were analyzed in each of three separate batches.

**Solid Mechanics Favors Developable Crystal Shapes**

To build towards an understanding of the vesicle size-based selectivity of crystal morphology, we start by considering the energetics of a vesicle having total area (*A*) and enclosed volume (*V*) where the fluid membrane contains a solid domain of fixed area fraction ($\phi$). While flexible vesicles need not be spherical, the area-integrated Gaussian curvature is constant and the standard (Helfrich model) bending energy of fluid membranes favors uniformly spherical vesicles,[13] both of which are frustrated by the presence of solid domain. To see this, we first estimate the energetics of forcing the solid crystal domain conform to a uniform spherical radius $R = \sqrt{A/4\pi}$. This energy is composed of the elastic strain energy imposed by the Gaussian curvature and the out-of-plane (i.e. mean curvature) bending energy cost. The former is proportional to $E(\text{strain}) \sim Y \phi^3 R^2$, where $Y$ is the 2D Youngs modulus of the solid, while the latter is proportional to the bending modulus $E(\text{bend}) \sim B\phi$.[33] For simplicity, bending modulus $B$ is taken to be the same for the fluid and solid phases. Since the modulus ratio $\sqrt{B/Y} = t$ gives a length scale, *t*, characterizing the elastic thickness,[10] and is thus of order of the nanometric lamella thickness, we expect the ratio $E(\text{bend})/E(\text{strain}) \sim (t/R)^2 \ll 1$. This implies that vesicles are likely to adopt large, mean-curvature bending deformations *without* imposing Gaussian curvature on solid domains. In other words, elasticity drives solid domains to take the form of zero-Gaussian *developable surfaces*[10, 38] either remaining planar or else bending cylindrically or in a locally conical geometry, while the $L_\alpha$ fluid phase will deform to accommodate all the non-zero Gaussian curvature of the closed vesicle.

These mechanics are borne out in the example observed vesicle shapes of Figure 3, selected with their solid crystalline domains oriented so they can be viewed side-on. Though the solids are not illuminated, the shape of the fluorescently-labeled $L_\alpha$ fluid membrane suggests a primarily flat or cylindrically curved solid domain shape. For instance in Figure 3A, the $L_\alpha$ fluid phase curves sharply at the boundary with a predominantly flat hexagonal inclusion. On the other hand, flower-shaped domains in Figure 3B appear to bend isometrically, with a planar core and largely cylindrical bending of the petals, strikingly similar to so-called "capillary origami" deformations of solid sheets on liquid drops.[39, 40] This suggests that the formation of highly non-convex and multi-lobed solid domains has similar effect in vesicles, to enhance the ability to conform closely to spherical shapes without in-plane strain. Unlike the capillary origami scenario, however, the energetics of the overall shape and morphology dependence is controlled by the overall bending energy of the fluid and solid phases.

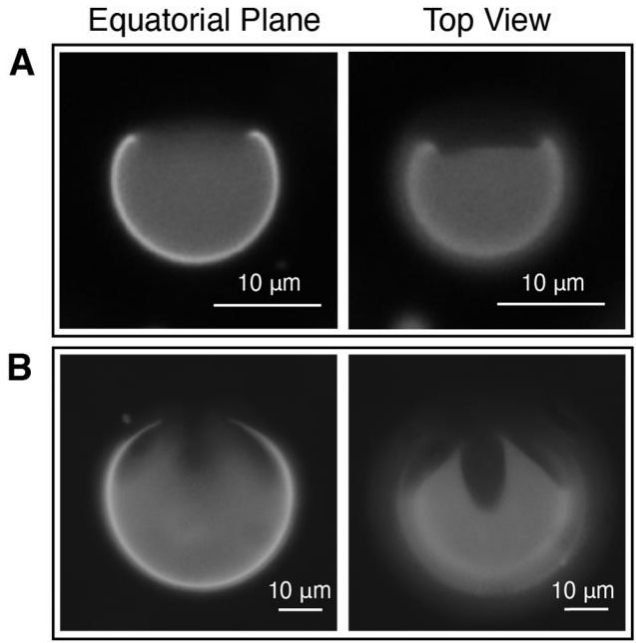

**Figure 3**. Micrographs showing the relationship between vesicle shape and solid domain bending (or lack of it) for typical solid domain shapes: **A**. hexagonal domain **B.** simple flower shaped domain. Equatorial and top views were obtained by changing the focus.

**Energetics of Inflation versus Crystal Shape**

We model the detailed distribution of elastic energy using Surface Evolver calculations[41] of closed vesicles having a single elastic solid (crystal) domain of area fraction, $\phi$, and a fixed dimensionless volume $\bar{v} = \sqrt{36\pi}\, V/A^{3/2} \leq 1$, under nearly inextensible conditions $t/R = 8.3\times10^{-5}$. $\bar{v}$ is a dimensionless measure of inflation, the actual volume normalized by the volume enclosed by an equal-area sphere. For the solid domain, we consider a range of flower-like shapes (detailed in the Methods section) which span from $\alpha =1$ for circular solid domains to $\alpha > 4$ for large petal morphologies. We first show elastic energy ground states in Figure 4A, for slightly inflated conditions (i.e. $\bar{v}$ above the minimal elastic energy state), mapping the distribution of both mean and Gaussian curvature over vesicle surfaces, resembling those in experiments. In all cases, it is found that the Gaussian curvature is expelled from the solid domains, confirming that their shapes closely follow developable surfaces, i.e. surfaces with everywhere zero Gaussian curvature. Second, we note for compact solid domains on inflated vesicles, large bending (mean curvature) is heavily concentrated in the fluid adjacent to the solid, to accommodate the transition from a mostly spherical fluid matrix and planar solid. Careful inspection of the pattern of mean-curvature in the solid shows that outer regions of the solid are folded along straight lines (known as the *generators* of developable shape), that do not intersect or end in the solid domain (see SI Figure S9), a necessary condition for avoiding Gaussian curvature.[10, 38] For $\alpha >1$, we observe that developable folding tends to concentrate at the bases of the petals, resulting in an effective planar flower core and overall a more uniform and energetically favorable mean-curvature distribution over the entire vesicle. Notably, the developable folding of the outer portions of solid domain shapes tends to focus Gaussian curvature in the fluid domain just outside solid domain, at the apparent intersection of cylindrical folding directions (i.e. generators). This folding is evident in the images in Figure 3B between the petals in real vesicles.

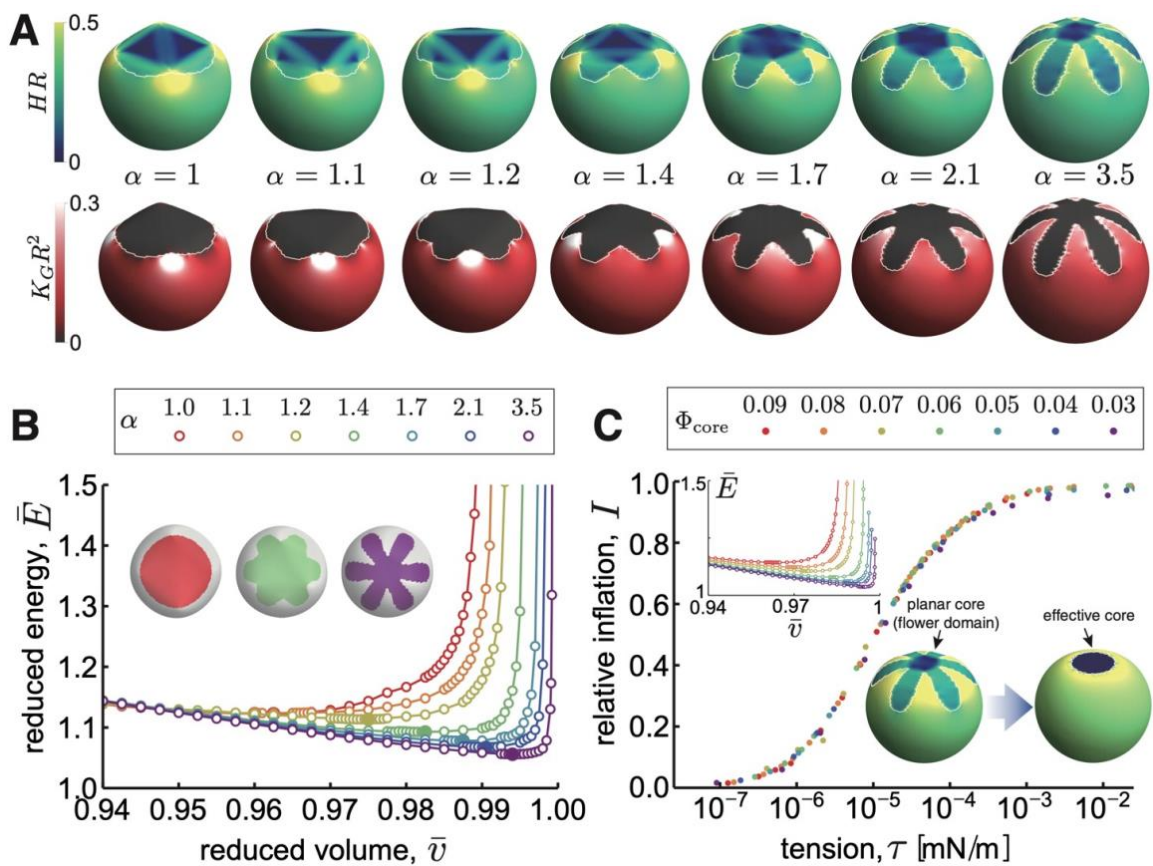

**Figure 4**. **A.** Simulated shapes of composite vesicles of solid area fraction ($\phi = 14\%$) with flower shapes of various petal/core ratios, showing mean ($H$) and Gaussian ($K_G$) curvature distributions. **B.** Reduced elastic energy of simulated vesicles versus reduced volume for the series of flower shapes in (A). The tension-free states ($\bar{v}_0$) are highlighted filled points on each curve. **C.** Comparison on an "effective core" model, in which the effect of the planar core in the flexible, flower-shaped solid is replaced by a rigid disk whose size is chosen to match the tension-free states for both models. The inset shows the energetics vs. reduced volume for the sequence of effective core values matching variable flower shapes in (B). Replotting in terms of the $I$ relative inflation (compared to maximal volume isoperimetric shapes) of effective core shapes shows a generic dependence of relative inflation on tension (here calculated for $R = 10$ μm shapes and taking $B = 25$ $k_BT$).

Figure 4B shows the reduced energy of the fluid-solid vesicles as a function of inflation, $\bar{v}$. In general, all domain shapes show a non-monotonic trend in energy, with a minimum at a particular inflation, which we denote as $\bar{v}_0$, that decreases over a series of solid shapes with increasingly larger petals (increasing $\alpha$). As shown in the Methods, the membrane tension $\tau$ is proportional to the slope of elastic energy versus $\bar{v}$. Hence the points $\bar{v}_0$ corresponding to vanishing tension, and $\bar{v} > \bar{v}_0$ are corresponding tensed and inflated shapes, the regime relevant to solid domain formation as we describe below. For a fixed solid area fraction, as is the case in experiments and computation, the elastic energy decreases with increasing degree of flowering ($\alpha$), and $\bar{v}_0$ shifts to higher inflation. Overall this shows that the larger the petal to core ratio the more uniform the overall mean curvature distribution that can be achieved by combination of the developable solid shape in the fluid matrix.

Figure 4C shows that the elastic energy versus inflation behavior of the more elaborately shaped flower domains over the range $\alpha \geq 1$ can be mapped onto a much simpler *effective core* model, where an area fraction $\Phi_{core}$ is modeled as rigid planar disk, with the remaining portion described by (mean curvature) bending elasticity only. The inset shows that the appropriate choice of effective core size mimics the zero-tension values $\bar{v}_0$ for the fixed solid fraction but variable petal shape, as shown in Figure S10. Notably, as shown in Figure S11 the effective core radius closely follows the variation of inner radii of flowers for the sequence $\alpha$ in Figure 4B. The effective core model provides a simple interpretation for the divergence of elastic energy at large inflation. A rigid circular core imposes a maximum inflation $\bar{v}^{max}$ shape, a spherical bulb joined to a planar core at finite angle, leading to diverging mean-curvature at the core edge. Using this maximal inflation we define $I = (\bar{v} - \bar{v}_0)/(\bar{v}^{max} - \bar{v}_0)$ as a measure of relative inflation, compared to the zero tension state at $\bar{v}_0$, and in Figure 4C analyze the relationship between relative inflation and tension, $\tau$, predicted by the effective core model. Notably, for all effective core sizes and a given $R$ (= 10 μm in this example), relative inflation follows the same dependence on $\tau$, switching from low- to high-relative inflation (where elastic energies grow significant) at a characteristic tension scale of $\sim 10^{-4}$ mN/m. As we describe below, the regulation of the inflation by tension is critical to the emergence of non-convex solid domain morphologies.

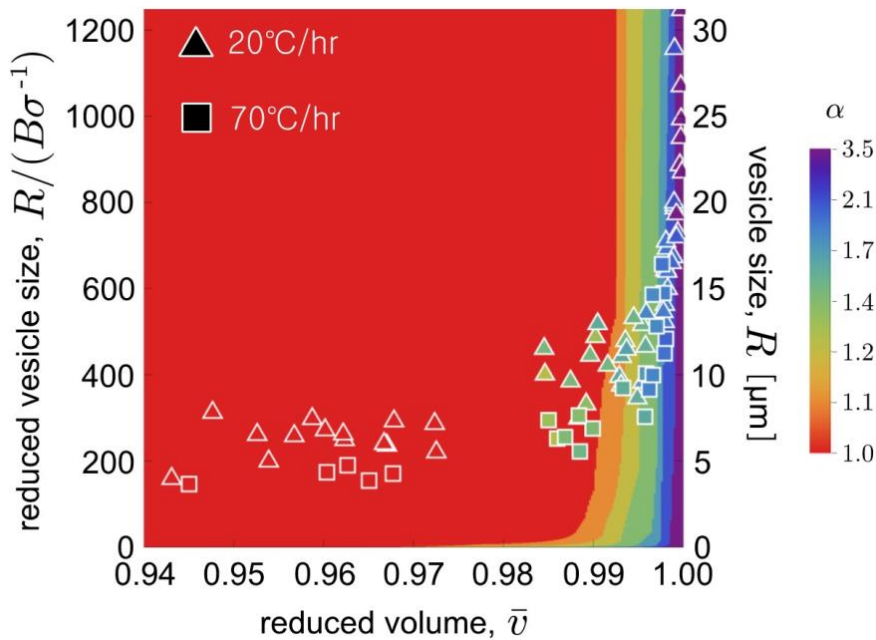

**Figure 5** Predicted state space of solid domain shapes (14% solid content) as function of vesicle inflation and size. Petal formation (i.e. $\alpha \geq 1$) becomes favorable when the relaxation of bending energy exceeds the line-energy ($\sigma$) cost for increasing the solid perimeter. The left axis shows predictions as function of vesicle radius scaled by $B/\sigma$, while the right axis shows the corresponding vesicle radius assuming an estimate $B/\sigma$ = 25 nm. Experimental results, with measured size $\alpha$ and $\bar{v}$ superposed on the model predictions, for two cooling rates: slow 20 °C/hour (triangles), and fast 70 °C/hour (squares).

While the elastic energy is decreased in vesicles whose solid domains contain longer petals that enable the solid to conform to more uniformly to curved quasi-spherical shapes (as shown in Figure 4B), increasingly elaborate flowers with greater fluid-solid perimeters incur an increasing cost of line energy. Comparing the line tension, $\sigma$, to the elastic cost of bending introduces a length scale $B/\sigma$ which is estimated to be in the range of 25 nm based on anticipated values of intra-vesicle fluid/solid domain boundaries. Figure 5 shows the resulting thermodynamically preferred domain shape as function of vesicle size and inflation. As line energy generically favors compact domains and, as the elastic energy differences for different petal ratios are relatively small at low inflation, compact domains are favored at low inflation, or smaller $\bar{v}$. In contrast, as inflation increases to values approaching spherical shapes $\bar{v} \to 1$, the elastic bending costs far exceed the additional line energies of petals, stabilizing petal formation. Indeed once full inflation is approached, the preference for flower shapes with large petals sets in abruptly and small changes in variables shift the preference for one shape flower over another. These results suggest that the predominant factor controlling flower stability and degree of petal to core area distribution is the degree of vesicle inflation, as captured by $\bar{v}$. Experimental data are included in Figure 5 for 2 different cooling rates, and show good qualitative agreement with modeling, most notably that the observed degree of flowering is strongly correlated with apparent increasing inflation, notwithstanding the increasing cost of line energy for flowers with large petals.

**Size-Dependent Tension and Solid Morphology in Thermally-Contracting Vesicles**

While competition between bending and line energy gives a preference for elaborate flower-shaped crystals on vesicles that are inflated and at elevated tensions, additional considerations explain why these elaborate solid domains grow selectively on large vesicles while compact domains grow on smaller vesicles. Figure 4C suggests that greater inflation and larger membrane tensions must be experienced by larger vesicles during nucleation and growth of solid domains.

Indeed, larger inflations and tensions on larger vesicles are expected when thermal contractions and permeability are considered. Vesicles cooled from the one phase region to nucleate and grow crystal domains experience thermal contractions, with the coefficient of thermal expansion, $\kappa \equiv \frac{1}{A}(\partial A/\partial T)_\tau$,[20] that tends to increase tension because the aqueous vesicle contents contract much less upon cooling than the predominantly hydrocarbon membrane.[42, 43] However, tension squeezes water from the vesicle, which tends to reduce tension itself. This transport process, occurring more slowly for large vesicles having a smaller surface/volume ratio, is governed by permeability $\mathcal{P}$, defined: $\partial V/\partial t = -A\,(2\tau/R)\,(\mathcal{P}/\rho)$, with $\rho$ the density of water. Note here that the pressure driving force for water transport is given by the Laplace term $2\tau/R$, providing the connection to membrane tension. Balancing the rate of tension increase due to thermal contractions with the rate of tension relaxation caused by water permeation reveals, for a constant rate of cooling, a steady state tension that grows with the cooling rate and square of vesicle radius,

$$\tau_{ss} \approx \left(\frac{\kappa \rho}{4\mathcal{P}}\right) |dT/dt|\, R^2 \qquad (1)$$

Worth noting, with annealing near 50°C in the one phase region, substantial cooling into the two-phase envelope preconditions tension before nucleation. This explains how the tendency for flower versus hexagon formation is predetermined before the crystals are visible and why domain shape is preserved during the growth occurring as temperature decreases further. Indeed growth occurs over less than ten additional degrees of cooling inside the two phase envelope, per Figure 1B.

For values of $\kappa = 0.005$ $K^{-1}$ [43] and $\mathcal{P} = 2.8 \times 10^{-16}$ s/μm,[44] the steady state tension can be as small as 0.5 mN/m for 10 μm diameter vesicles and as large as 18 mN/m for 60 μm diameter vesicles cooled at 0.3 °C/min, as done here. The latter exceeds the lysis tension, which falls in the range 7-10 mN/m.[44] Also the flower-shaped domains may be on vesicles experiencing lower tensions than required for stripe solid domain formation, a different polymorph.[17, 45] Thus if large vesicles are not destroyed, they undergo burst-reseal processes that maintain their membrane tensions near or below lysis conditions.[46] Indeed, micropipette aspiration after cooling and transferring vesicles to an open chamber (a process allowing some tension decay) reveals near zero tension in small vesicles for instance with diameters less than 15 μm, and substantial tension for large vesicles for instance with diameters exceeding 25 μm, in Figure 6.

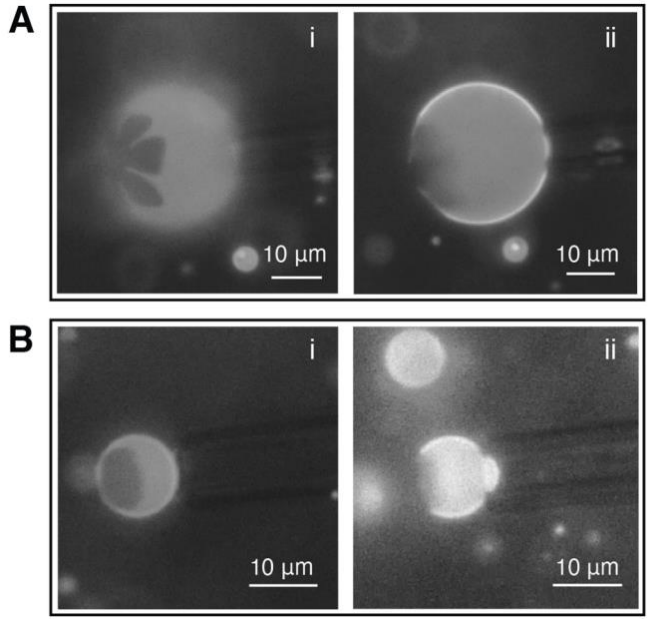

**Figure 6.** Micrographs of typical A) flower-containing and B) hexagon containing vesicles, (i) before and (ii) after aspiration into a micropipette. The focal plane before aspiration (i) shows the crystal pattern on each vesicle. After aspiration in (ii), the focal plane was adjusted to the equatorial plane of each vesicle to enable viewing any membrane projection into the pipette. In Aii, the suction was relatively strong, 13.45 cm H$_2$O (1.32 kPa), still not producing a projection and indicating high membrane tension. In Bii, a relatively low suction of 0.26 cm H$_2$O (0.025 kPa) drew the membrane into the micropipette, indicating its negligible tension compared with the vesicle in A.

Finally we note that the growth of steady state tension as $|dT/dt| R^2$ from estimate in eq. (1) suggests that the same morphology transitions – from compact to non-convex flower domains – can be achieved at a *different size range* simply through a change in *cooling rate*. Indeed this shifting of the transition from hexagons to flowers was seen, in experiments, to occur on smaller vesicles when the cooling rate was increased. This behavior, comparing two cooling rates and vesicles of different sizes is included in Figure 5. For example, comparing similarly sized vesicles (along a horizontal cut). Those vesicles cooled more quickly reside to the right, with greater inflations and higher alpha values, indicative of more elaborate flowers.

**Outlook**

The work illustrates how the morphologies of 2D crystals growing in an elastic 2D fluid are controlled by a competition between bending energy and line tension in a manner that inverts the expected crystalline domain morphology dependence on curvature radius understood to control crystallization on spherical templates. Additional properties of the 2D fluid, its coefficient of thermal expansion and permeability, allow for tension and therefore bending energy to be systematically controlled, leading to scalable manipulation of 2D crystal morphology in vesicle of different sizes. The current study suggests further scaling that could produce at least thousands of such 2D crystals of controlled morphology in individual batches, amenable to separation based on vesicle size, or density gradients. Thus we demonstrate how 2D membrane fluids such as phospholipids can be implemented to controllably produce a targeted 2D crystalline morphologies based on their physical properties and select these morphologies without changing chemical composition.

**METHODS**

**GUV Electroformation and Phase Separation**

1,2-dioleoyl-sn-glycero-3-phosphocholine (DOPC), 1,2-dipalmitoyl-sn-glycero-3-phosphocholine (DPPC), and fluorescent tracer lipid, 1,2-dioleoyl-sn-glycero-3-phosphoethanolamine-N-(lissamine rhodamine B sulfonyl) (ammonium salt) (Rh-DOPE) were purchased from Avanti Polar Lipids (Alabaster, AL). Vesicles having a 30/70 weight ratio (31/69 molar ratio) DPPC/DOPC plus 0.1 mol% Rh-DOPE, were electroformed on platinum (Pt) wires by established methods but with the sucrose preheated and the electroforming temperature maintained in the range 55-70°C to ensure vesicles were formed in the one phase region of the phase diagram, all having the same membrane composition. After electroformation, the stock vesicle suspension was harvested in a syringe and allowed to cool to room temperature for later use, within 2-3 days.

Studies of solid domain formation employed a 10-fold dilution of the stock vesicle solution in deionized (DI) water, which was transferred to a 10 mm x 10 mm closed chamber made from two coverslips and parafilm spacers. The chamber height varied in the range 0.1-0.5 mm depending on the desired solution volume (10-50 μL) for a specific experiment. The chamber was mounted on a custom-built temperature-control stage, heated to 55 °C for 5 minutes, cooled at 70 °C/hr to 42°C, and then cooled at 20 or 70 °C/hr to room temperature, comprising the regular or fast cooling rates, respectively.

Vesicle and crystalline domain shapes were observed using a Nikon Eclipse TE 300 inverted epifluorescence microscope equipped with a 40X long working distance air fluorescence objective. Images were recorded with a pco.panda 4.2 sCMOS monochrome camera with a resolution of 0.17 μm/pixel at 40X, and analyzed using Nikon NIS Elements imaging software.

**Micropipette Aspiration**

Micropipette aspiration employed micropipettes pulled from glass capillaries to produce straight micropipette tips having inner diameters in the range 3-10 μm and flat ends. They were passivated with adsorbed bovine serum albumin before use. These were attached to a suction manometer equipped with a Validyne pressure transducer (model CD223) to control aspiration pressure and record values in the Nikon Elements Software. Micropipette experiments were conducted in an open-sided chamber consisting of two coverslips, passivated with adsorbed albumin, spaced by a microscope slide. After controlled cooling in the closed chamber to produce one solid crystal per vesicle, the liquid vesicle suspension was transferred to the open chamber[47] where it was held in place through capillary forces during micropipette studies which were conducted within 10-20 minutes after the end of the cooling in the close chamber. Membrane tension was determined employing the Laplace equation, as previously detailed.[20]

**Thermodynamic modeling of solid-fluid composite vesicles**

Our model considered vesicles of fixed total area $A$ and enclosed volume $V$, composed of area fraction $\phi$ of 2D solid and the remaining $(1 - \phi)$ is fluid membrane. The solid domain is modeled as an elastic plate with 2D Youngs modulus $Y$, Poisson ratio $\nu = 0.4$ and plate bending modulus $B$

$$E_{\text{solid}} = \frac{Y}{2(1+\nu)} \int_{\text{solid}} dA \left[ (\text{Tr } \varepsilon)^2 + \frac{\nu}{1-\nu} \text{Tr } \varepsilon^2 \right] + \frac{B}{2} \int_{\text{solid}} (2H)^2 \qquad (2)$$

where $\varepsilon$ is the 2D strain and $H$ is the mean curvature. We consider the Helfrich model of the (fixed area) fluid region, with a bending energy $E_{\text{fluid}} = \frac{B}{2} \int_{\text{fluid}} (2H)^2$ where we assume the same bending modulus as the solid phase for simplicity. Note that, because the 2D solid expels Gaussian curvature, bending terms coupling to Gaussian curvature in both fluid and solid phases are shape independent.

We use Surface Evolver to minimize eq. (2) subject to constraints of fixed solid and fluid phase area, and enclosed volume, and define reduced elastic energy $\bar{E} \equiv \frac{E_{\text{solid}} + E_{\text{fluid}}}{8\pi B}$ as the ratio of elastic energy of the composite relative to spherical fluid vesicle of the same bending stiffness.

We consider 6-fold symmetric solid domain shapes with variable petal to core aspect ratios. The shape of the solid domain is defined it perimeter in a planar reference state. We model these shapes by the following family of radial functions:

$$r(\theta) = r_0 + \frac{1}{2} a r_0 \cos(6\theta) - \frac{1}{10} a r_0 \cos(12\theta) \qquad (3)$$

where petal length is controlled through parameter $a$ (i.e. $\alpha = r(0)/r(\pi/6) = (5 + 2a)/(5 - 3a)$) and $r_0$ can be adjusted to fix the solid domain area. We describe the procedure for initializing closed meshes with solid domains of these shapes, subsequent energy minimization in Surface Evolver and characterization in the Supporting Information.

For the *effective core* simulations, we hold a circular domain on the vesicle of a fixed area fraction $\Phi_{\text{core}}$ to maintain a planar configuration. To determine the effective core for a given $\alpha$, $\bar{E}(\bar{v})$ is computed for effective core model and $\Phi_{\text{core}}$ is adjusted so that minimal energy occurs for the same inflation $\bar{v}_0$ as the 14% solid with that $\alpha$ (see Supporting Information).

The *tension* $\tau$ in the membrane can be related from the reduced energy $\bar{E}(\bar{v})$ dependence on inflation through the thermodynamic relation

$$\tau = -\left(\frac{\partial E}{\partial A}\right)_{V,\phi} = -8\pi B \left(\frac{\partial \bar{E}}{\partial \bar{v}}\right)_\phi \left(\frac{\partial \bar{v}}{\partial A}\right)_V = \frac{12\pi B}{A} \bar{v} \left(\frac{\partial \bar{E}}{\partial \bar{v}}\right)_\phi \qquad (4)$$

To access the thermodynamic stability of flowered shapes, we consider the total energy

$$E = E_{\text{solid}} + E_{\text{fluid}} + \sigma P \qquad (5)$$

where $\sigma$ are the line energy and perimeter of the boundary edge between the solid and fluid domain. The state map in Figure 5 is determined by finding the value of $\alpha$ that minimizes the total energy in eq. (5). When $t/R \to 0$ the elastic energy is strictly strain free and derives from (mean-curvature) bending only. As in the case of fluid vesicles, this isometric limit is independent of vesicle *size*, depending only on $B$ and dimensionless ratios, $\bar{v}$, $\phi$ and $\alpha$. The line energy, however, is proportional to vesicle (and solid domain) size. Hence, as noted previously comparing elastic and line energies requires comparing the vesicle radius to a the length scale $B/\sigma$.[42, 48, 49] For mapping to the state space in Figure 5, we take for the bending modulus $B = 25\ k_BT$ (corresponding the fluid phase stiffness) and an estimate for line energy $\sigma = 1\ k_BT/\text{nm}$.

*Acknowledgements*: This work was supported by DOE DE-SC0017870.

The authors declare they have no competing interests.

*Author contributions:* M.M.S. conceived and oversaw the experimental program and the thermal-contraction /membrane permeation mechanism, performed calculations of tension evolution and steady state, developed data analysis strategies, and wrote most of the paper. G.M.G. conceived and oversaw modeling of membrane mechanics, developed figures, and wrote part of the paper. H.W. executed the majority of experiments and data analysis, developed the experimental figures with W. X. conducting and analyzing micropipette experiments. G.J. executed calculations of membrane mechanics and developed figures based on modeling.

*Supplemental Information:*

**Flowering of Developable 2D Crystal Shapes in Closed, Fluid Membranes**

Hao Wan,[1] Geunwoong Jeon,[2] Weiyue Xin,[3] Gregory M. Grason,[1] and Maria M. Santore[1,*]

1. Department of Polymer Science and Engineering, University of Massachusetts, 120 Governors Drive, Amherst, MA 01003
2. Department of Physics, University of Massachusetts, 710 N. Pleasant Street, Amherst, MA 01003
3. Department of Chemical Engineering, University of Massachusetts, 686 N. Pleasant Street, Amherst, MA 01003

# Table of contents

# 1. Estimating solid area fraction from the phase diagram

Here, we use simple mass balance (lever arm rule) to estimate the solid area fraction from the phase diagram for the two-component DOPC/DPPC lipid vesicle membrane. A detailed derivation of the solid area fraction is found in the supporting information of a work by Chen and Santore.[1]

The solid area fraction $\phi$, of the vesicle surface containing solid domains, at equilibrium, is

$$\phi = \left(1 + \frac{1}{R}\frac{z_A - y_A}{x_A - z_A}\right)^{-1} \qquad (1.1)$$

Here $R \equiv \underline{A}_S/\underline{A}_L$ is the ratio of the lipid molar area in the solid phase, $\underline{A}_S$, to that in the fluid phase, $\underline{A}_L$. R depends on the particular solid and fluid and is temperature dependent.

For a phase-separated vesicle at room temperature, $x_A$, $y_A$ are the DPPC mole fractions in the fluid and solid phases, respectively. $z_A$ is the overall DPPC mole fraction.

The molecular lipid area in the fluid phase is estimated as the molecular area of DOPC in the fluid phase since DOPC is the fluid phase component ($\underline{A}_L \approx A^F_{DOPC}(@\ 22\ °C)$). Nagle and Tristram-Nagle[2] report the area per lipid molecule for DOPC in the fluid $L_\alpha$ phase as $A^F_{DOPC}(@\ 30°C) = 72.5\ Å^2$. The area for DOPC at room temperature is then adjusted using the area thermal expansivity $\kappa = (1/A)(\partial A/\partial T)_\tau$, with a value of 0.003/°C based on similar treatment from Nagle and Tristram-Nagel.[2] This gives $A^F_{DOPC}(@\ 22°C) = A^F_{DOPC}(@\ 30°C)\ e^{\kappa\Delta T} = 70.8\ Å^2$.

The solid phase molecular lipid area is estimated as the molecular area of DPPC in the ripple $P_{\beta'}$ phase at 22 °C ($\underline{A}_S \approx A^S_{DPPC}(@\ 22\ °C)$) since solid domains have nearly pure DPPC content. Nagle and Tristram-Nagle[2] report the area per DPPC molecule in the fluid phase as

$A_{\text{DPPC}}^{\text{F}}(@\ 50°C) = 64.0\ \text{Å}^2$. The extrapolated molecular area for DPPC at 22°C ($A_{\text{DPPC}}^{\text{S}}(@\ 22°C)$) is calculated using $\kappa = 0.003 - 0.006$ /°C from 50 to 42 °C for fluid $L_\alpha$ phase, $\kappa = 0.003 - 0.006$ /°C from 41 to 22°C for rippled $P_{\beta'}$ solid phase, and an approximately total 17% areal reduction from the fluid $L_\alpha$ to the rippled $P_{\beta'}$ solid phase through the main transition at 41~ 42°C.[3] This gives $A_{\text{DPPC}}^{\text{S}}(@\ 22°C) = 45.2 - 49.0\ \text{Å}^2$.

According to the phase diagram, at room temperature (22 °C), $x_A$ is $0.17 \pm 0.02$ and $y_A$ is approximated as 0.95 since phase separation in DOPC/DPPC mixtures is known to produce solid domains that are nearly pure in DPPC.[4, 5, 6, 7, 8] $z_A$ is 0.314 (30wt% DPPC).

Then, from Equation (1.1), solid area fraction $\phi$ is estimated to be $11 - 15$ %. The lower and upper limit are estimated using $A_{\text{DPPC}}^{\text{S}}(@\ 22°C)$ value of 45.2 Å$^2$ and 49.0 Å$^2$, respectively.

If the solid phase contains pure DPPC ($y_A=1$), from Equation (1.1), solid area fraction $\phi$ is estimated to be $10 - 15$ %.

## 2. Solid area fraction calculation from vesicle images

*2A. Solid area fraction for hexagonal domains*

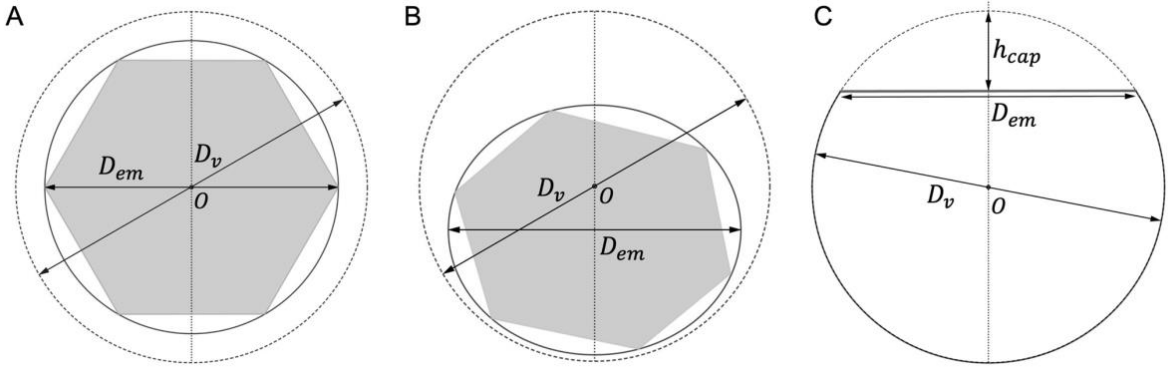

**Figure S1**. Projections of a hexagonal shape on the flat plane of a truncated sphere, viewed from three different angles. (A) Perpendicular view demonstrating the hexagon in its original shape. (B) An oblique view resulting in a distorted representation of the hexagon, which resembles a typical micrograph from fluorescent microscopy. (C) Side view

To simplify the estimation of the solid area fraction of a hexagonal domain, the solid hexagonal domain and the fluid part inside the hexagon's circumscribed circle is treated as a plane while the other part of the fluid vesicle is treated as a section of a sphere, as shown in Figure S1C. To calculate the solid area fraction of a hexagonal domain, the following quantities are measured, including:

$D_v$: Vesicle Diameter

$D_{em}$: Length of the major axis of the ellipse that goes across all the vertices of the deformed hexagon (as shown in Figure S1B). This diameter is irrelevant to the position of the hexagon on the vesicle, as shown in Figure S1A and S1B.

The solid area fraction of the hexagon $\phi_{hex}$ is then calculated as:

$$\phi_{\text{hex}} = \frac{3\sqrt{2}}{2\pi} \times \frac{\pi \left(\frac{D_{\text{em}}}{2}\right)^2}{4\pi \left(\frac{D_{\text{v}}}{2}\right)^2 - 2\pi \left(\frac{D_{\text{v}}}{2}\right) h_{\text{cap}} + \pi \left(\frac{D_{\text{em}}}{2}\right)^2} \quad (2.1)$$

where

$$h_{\text{cap}} = \frac{D_v}{2} - \sqrt{\left(\frac{D_v}{2}\right)^2 - \left(\frac{D_{\text{em}}}{2}\right)^2} \qquad (2.2)$$

*2B. Solid area fraction for flower-shaped domains*

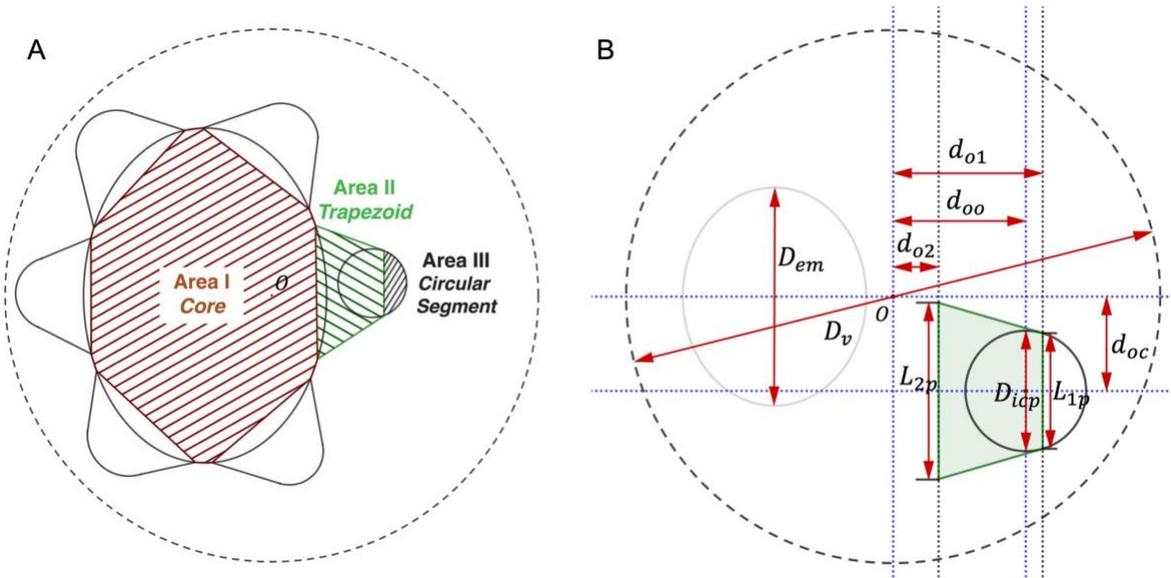

**Figure S2**. (A) Projection of a ninja star flower shape on a truncated sphere, with a hexagonal core on the flat plane and the petals taking cylindrical curvature, viewed from an oblique angle, which resembles a typical micrograph from fluorescent microscopy. The shaded regions show the decomposition of the flower shape to facilitate solid area calculation. (B) Scheme of all the quantities measured for calculating the solid area fraction of a flower domain. The positions and sizes of the core and petal are altered to facilitate the view of those quantities and may not resemble a real micrograph.

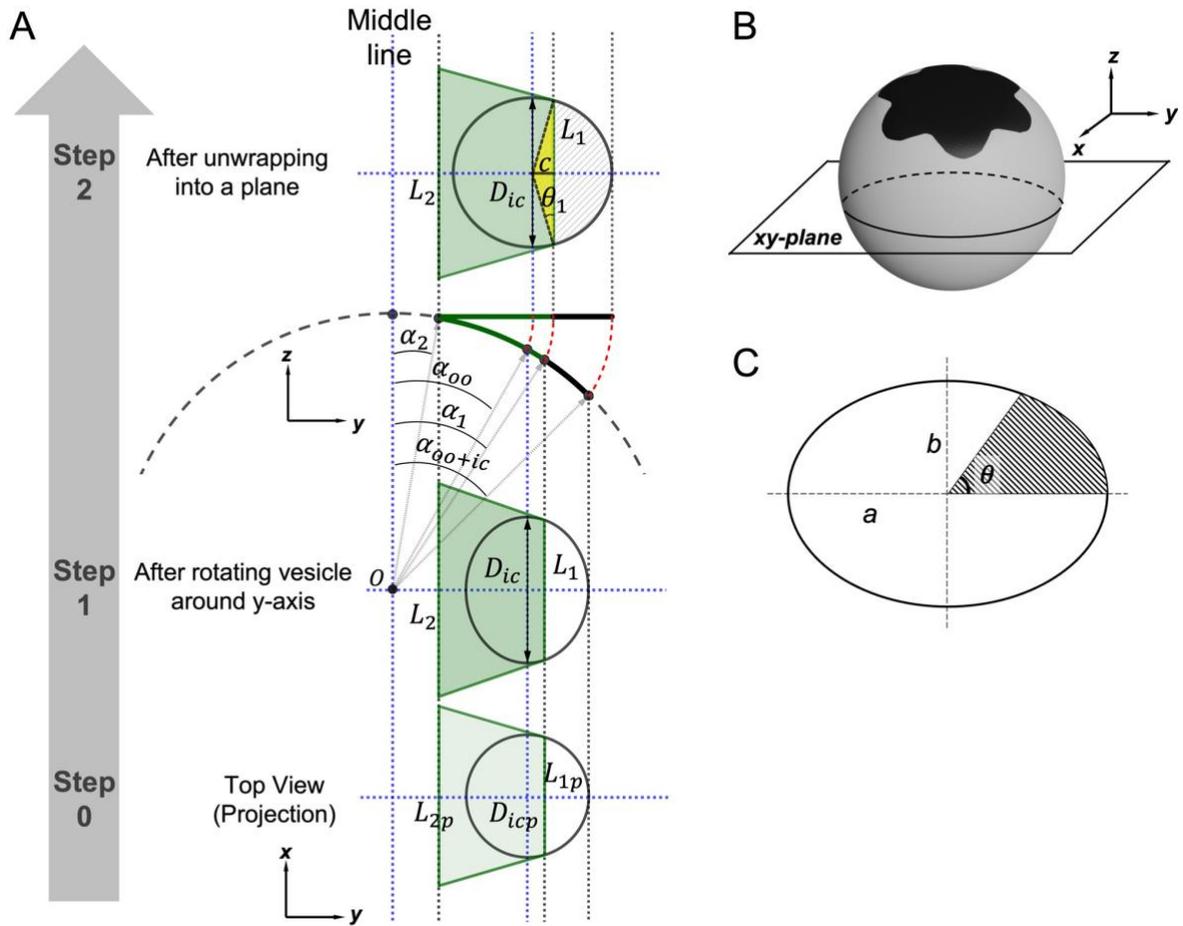

**Figure S3**. Illustration for the area calculation of a flower petal. (A) Procedures for transforming the measured quantities on a sphere in the projection view to a planar view. (B) Scheme for a perspective of a vesicle under fluorescent microscopy. A micrograph is taken by moving along the z-axis until a preferred focus plane is chosen. Vesicles with solid domains facing the z-direction are chosen to facilitate area measurement. (C) Scheme for a section (the shaded region) of an ellipse. a, b and θ are the major axis, minor axis, and the polar angle of the ellipse.

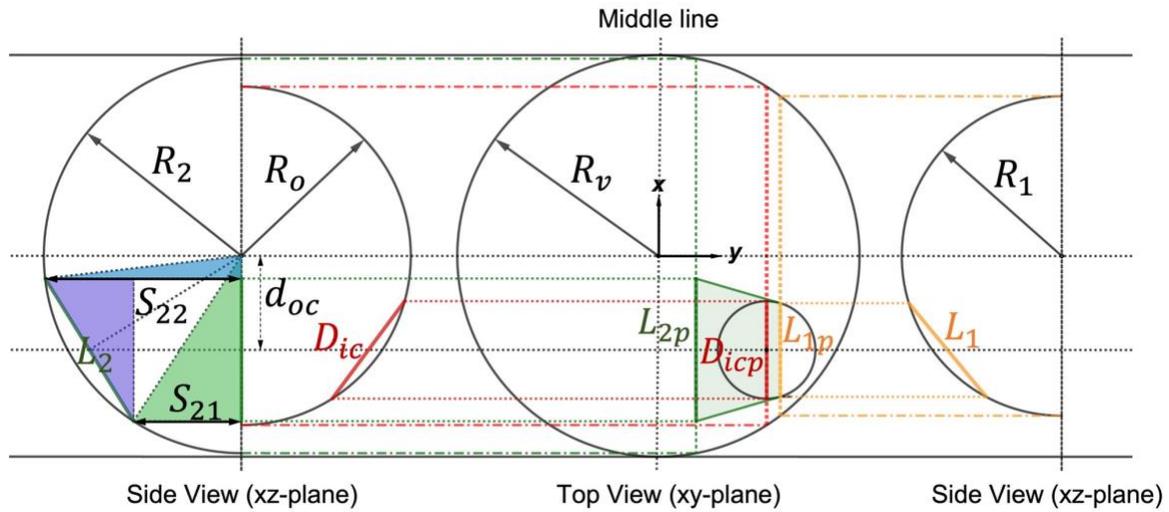

**Figure S4**. Scheme for calculating the real length of the two bases of the trapezoid (green and red) and the diameter of the inscribed circle (yellow). The projection length(xy-plane) for those quantities corresponds to the chord lengths in corresponding xz-planes of the sphere.

To simplify the measurement of the solid area fraction of a flower domain (we use ninja star flower as an example here), the flower domain is assumed to have six-fold symmetry and be composed of a hexagon core (Area I) which is flat and six petals which have cylindrical curvature. The other fluid part of the vesicle is assumed to have identical spherical curvature, ignoring the complexity of the curvature for the connection region between the solid petals contour and the fluid membrane. For the purpose of measuring solid area fraction, the error for this simplified treatment should be small and not enough to influence the result. Each petal is treated as the combination of a trapezoid (Area II) and a section of a circle (Area III) whose center is on the bisector of the trapezoid and the circle is tangent to both vertices of the short base of the trapezoid (Figure S2A). The circle will be named the "inscribed circle" of the trapezoid from now on for simplification.

To measure the solid area fraction of a flower domain, the following quantities (9 in total) are measured, as shown in Figure S2B (the position of the petal and the core in Figure S2B is just for a clear illustration of the quantities and might be different from a real measurement):

    $D_v$: Vesicle Diameter

$D_{em}$: Length of the major axis of the ellipse that go across all the six grooves between the petals.

$D_{icp}$: Projection length of the diameter of the inscribed circle of the trapezoid.

$L_{1p}$: Projection length of the short base of the trapezoid

$L_{2p}$: Projection length of the long base of the trapezoid

$d_{o1}$: Perpendicular distance between the center of the equatorial circle (Point O) and the line of $L_{1p}$

$d_{o2}$: Perpendicular distance of point O to the line of $L_{2p}$

$d_{oo}$: Perpendicular distance of point O to the bisector of the inscribed circle of the trapezoid that is parallel to the trapezoid base.

$d_{oc}$: Perpendicular distance of point O to the bisector of the trapezoid

Then, area I, II and III are calculated as followed:

Area I:

The hexagonal shape in Area I is treated roughly as the inscribed hexagon of the ellipse, although it might not be exactly a perfect hexagon.

$$A_I = \frac{3\sqrt{2}}{2\pi} \times \pi \left(\frac{D_{em}}{2}\right)^2 \tag{2.3}$$

Area II and Area III:

Before showing the equations for calculating Area II and III, we will first introduce the basic idea of the calculation here. The long base of the trapezoid is connected to the side of the hexagonal shape of Area I. To measure the real area of the petal unwrapped in a plane, we can consider in two steps, as shown in Figure S3A. Since the petal is assumed to have cylindrical curvature, the two bases of the trapezoid are straight lines. A fluorescent image in our experiment is a projection view image in the xy-plane from a certain in-focus depth in the z-direction (as shown in Figure S3B). We can first imagine rotating the vesicle around y-axis to position the petal to the center of the view (Figure S3A Step 0 to Step 1). The real length of the two bases ($L_1$ and $L_2$) are chord lengths in corresponding xz-plane slice of the vesicle, as shown in Figure S4. Next, we can imagine unwrapping the petal onto a plane, as shown in Figure S3A Step 1 to Step 2, which will elongate features in the bisector direction of the trapezoid. We assume the inscribed circle has an ellipse shape after unwrapping (not sure about the major and minor axis).

In the first step, the length of the two trapezoid bases $L_1$, $L_2$ and the diameter of the inscribed circle $D_{ic}$ (now is an ellipse rather than a circle) are calculated. Here, the calculation of $L_2$ is used as an example (see Figure S4 green lines and Equation (2.4) - (2.7)) and the calculation of $L_1$ and $D_{ic}$ are similar and omitted here.

$$d_{o2}^2 + R_2^2 = \left(\frac{D_v}{2}\right)^2 \tag{2.4}$$

$$S_{21}^2 + \left(\frac{L_{2p}}{2} + d_{oc}\right)^2 = R_2^2 \text{ (Green Triangle)} \tag{2.5}$$

$$S_{22}^2 + \left(\frac{L_2'}{2} - d_{oc}\right)^2 = R_2^2 \text{ (Blue Triangle)} \tag{2.6}$$

$$L_{2p}^2 + (S_{22} - S_{21})^2 = L_2^2 \text{ (Purple Triangle)} \tag{2.7}$$

Next, the height of the trapezoid ($H_{tpz}$) is calculated as

$$H_{tpz} = \begin{cases} R_v(\alpha_1 - \alpha_2), & \text{if trapezoid on one side of the middle line in Figure S3A} \\ R_v(\alpha_1 + \alpha_2), & \text{if the trapezoid goes across the middle line, not shown here} \end{cases} \tag{2.8}$$

$$\alpha_1 = \arcsin\frac{d_{o1}}{R_v} \tag{2.9}$$

$$\alpha_2 = \arcsin\frac{d_{o2}}{R_v} \tag{2.10}$$

Then, the area of the trapezoid (Area II) is:

$$A_{II} = (L_1 + L_2) \times \frac{H_{tpz}}{2} \tag{2.11}$$

The area of the ellipse segment (Area III) is the difference between the area of unwrapped ellipse section and the small triangle (as shown in Figure S3A). To calculate area III, the following quantities need to be introduced, including the major (a) and minor axis (b) of the unwrapped ellipse, the height of the triangle (c), and the angle $\theta_1$, as shown in Figure S3A.

As shown in Figure S3C, for a section of an ellipse, the area is

$$A_{\text{elp section}} = \frac{ab}{2} \times \arctan\left(\frac{a}{b}\tan\theta\right) \tag{2.12}$$

For the case in Figure S3A, $R_{ic}$ ($D_{ic}/2$) is either the major or minor axis, and $R'_{ic}$ is defined as the other axis. $R'_{ic}$ is calculated as followed:

$$R'_{ic} = R_v \times (\alpha_{oo+ic} - \alpha_{oo}) \tag{2.13}$$

$$\alpha_{oo} = \arcsin\frac{d_{oo}}{R_v} \tag{2.14}$$

$$\alpha_{oo+ic} = \arcsin\frac{d_{oo} + R_{ic}}{R_v} \tag{2.15}$$

The height of the triangle (c) is:

$$c = R_v \times (\alpha_1 - \alpha_{oo}) \tag{2.16}$$

The angle $\theta_1$ is:

$$\theta_1 = \arctan\frac{2c}{L_1} \tag{2.17}$$

After substitute a, b and $\theta$ in Equation (2.12) for corresponding quantities for the case in Figure S3A, we get

$$A'_{\text{elp section}} = \begin{cases} R_{ic}R'_{ic} \times \arctan\left(\frac{R'_{ic}}{R_{ic}}\cot\theta_1\right) & \text{if } R_{ic} < R'_{ic} \\ R_{ic}R'_{ic} \times \left(\frac{\pi}{2} - \arctan\left(\frac{R_{ic}}{R'_{ic}}\tan\theta_1\right)\right) & \text{if } R_{ic} > R'_{ic} \end{cases} \tag{2.18}$$

The area of the small triangle is:

$$A_{\text{triangle}} = \frac{cL_1}{2} \tag{2.19}$$

Then, Area III is:

$$A_{III} = A'_{\text{elp section}} - A_{\text{triangle}} \tag{2.20}$$

The area of a flower is:

$$A_{\text{flower}} = A_I + 6(A_{II} + A_{III}) \tag{2.21}$$

The solid area fraction of the flower domain is:

$$\phi_{\text{flower}} = \frac{A_{\text{flower}}}{4\pi\left(\frac{D_v}{2}\right)^2 - 2\pi\left(\frac{D_v}{2}\right)h_{\text{cap}} + \pi\left(\frac{D_{\text{em}}}{2}\right)^2} \quad (2.22)$$

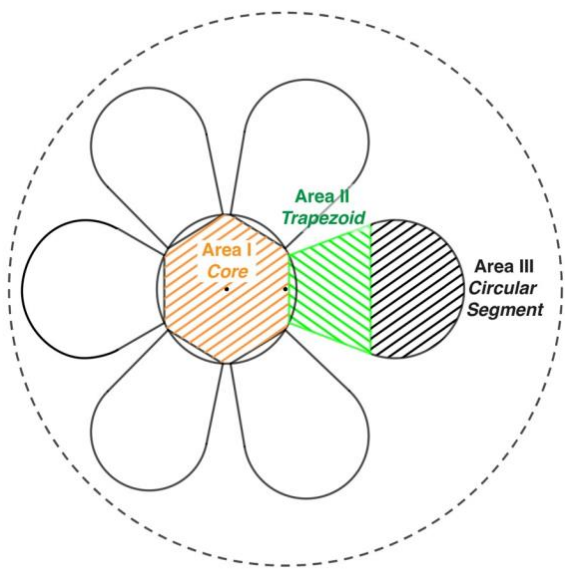

**Figure S5**. Projection of a simple flower shape on a truncated sphere, with a hexagonal core on the flat plane and the petals taking cylindrical curvature, viewed from an oblique angle, which resembles a typical micrograph from fluorescent microscopy. The shaded regions show the decomposition of the flower shape to facilitate solid area calculation, which resembles the case of ninja star flower.

In the previous case, ninja star flower was used as an example for solid area fraction calculation. For a simple flower as shown in Figure S5, the shapes can be divided to three parts as well, including the hexagonal shape core (Area I), the trapezoid (Area II), and the circular segment (Area III). The difference is that the "inscribed circle" of the trapezoid here is tangent to both vertices of the long base of the trapezoid. As a result, the procedure for calculating the solid area is almost the same as the previous case, and only minor revisions are needed. Here, only the equations that differ from the previous case are shown and all other equations are the same as before.

$$H_{\text{tpz}} = \begin{cases} R_v|\alpha_1 - \alpha_2|, & \text{if the trapezoid is on one side of the middle line} \\ R_v(\alpha_1 + \alpha_2), & \text{if the trapezoid go across the middle line} \end{cases} \quad (2.8')$$

$$R'_{ic} = \begin{cases} R_v(\alpha_{ic-oo} + \alpha_{oo}), & \text{if O is between center of the inscribed circle and petal edge} \\ R_v(\alpha_{oo+ic} - \alpha_{oo}), & \text{for almost all other circumstances} \end{cases} \quad (2.13')$$

Where,

$$\alpha_{ic-oo} = \frac{\arcsin(R_{ic} - d_{oo})}{R_v} \quad (2.15')$$

$$c = R_v |\alpha_2 - \alpha_{oo}| \quad (2.16')$$

$$\theta_1 = \arctan \frac{2c}{L_2} \quad (2.17')$$

$$A'_{elp\ section} = \begin{cases} \pi R'_{ic} R_{ic} - R_{ic} R'_{ic} \times \arctan\left(\frac{R'_{ic}}{R_{ic}} \cot \theta_1\right) & \text{if } R_{ic} < R'_{ic} \\ \frac{\pi}{2} R'_{ic} R_{ic} + R_{ic} R'_{ic} \times \arctan\left(\frac{R_{ic}}{R'_{ic}} \tan \theta_1\right) & \text{if } R_{ic} > R'_{ic} \end{cases} \quad (2.18')$$

$$A_{triangle} = \frac{cL_2}{2} \quad (2.19')$$

$$A_{III} = A'_{elp\ section} + A_{triangle} \quad (2.20')$$

*2C. Images of solid area fraction measurement*

Based on the methods mentioned above, ~ 15 vesicles for each shape were measured for solid area fraction. The original images and their corresponding solid area fraction ($\phi$) are shown in Figure S6-S8 for hexagonal, ninja star flower, and simple flower domain respectively.

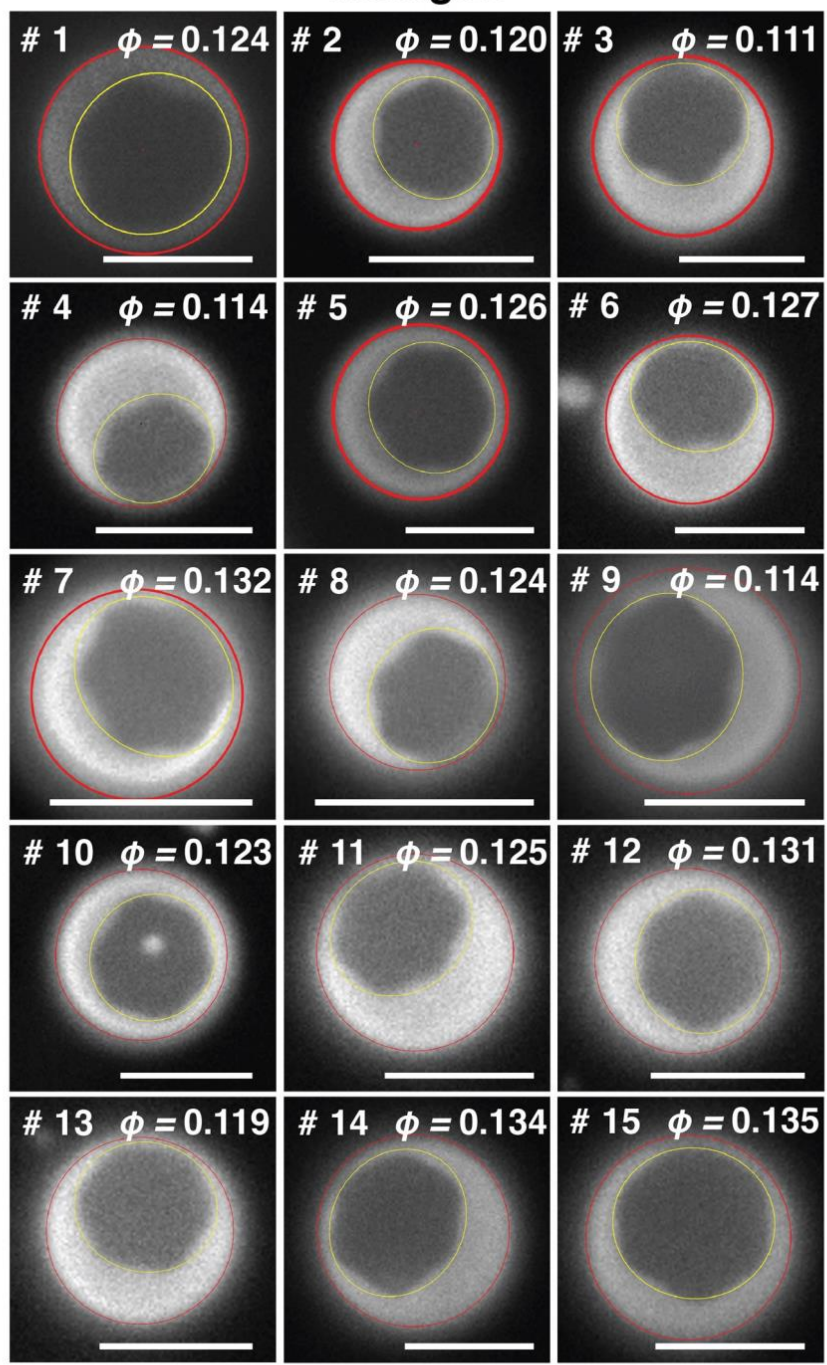

All Scale Bar: 10 μm

**Figure S6**. Fluorescent images and corresponding solid area fraction for 15 different vesicles with hexagonal domains. Embedded lines in each micrograph show the measured quantities based on the previously described methods.

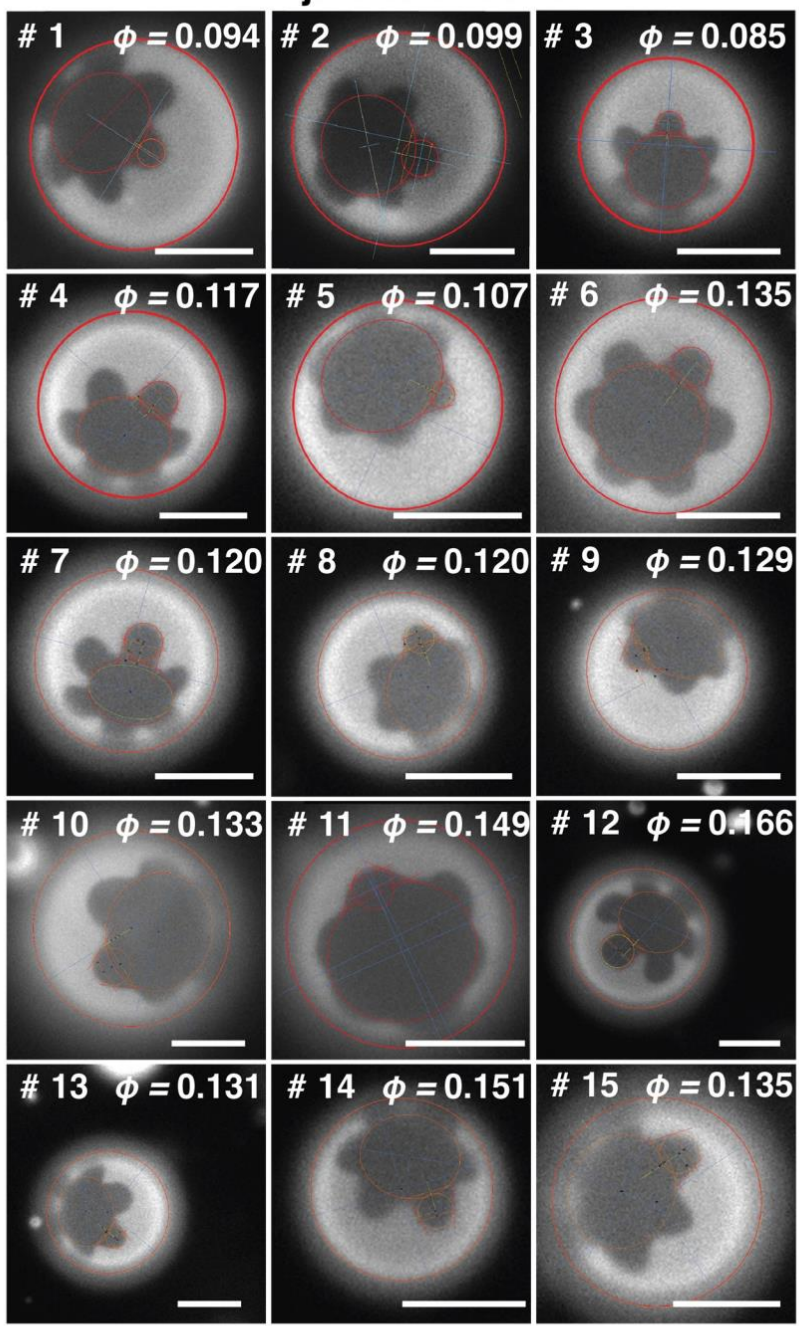

**Figure S7.** Fluorescent images and corresponding solid area fraction for 15 different vesicles with ninja star flower domains. Embedded lines in each micrograph show the measured quantities based on the previously described methods.

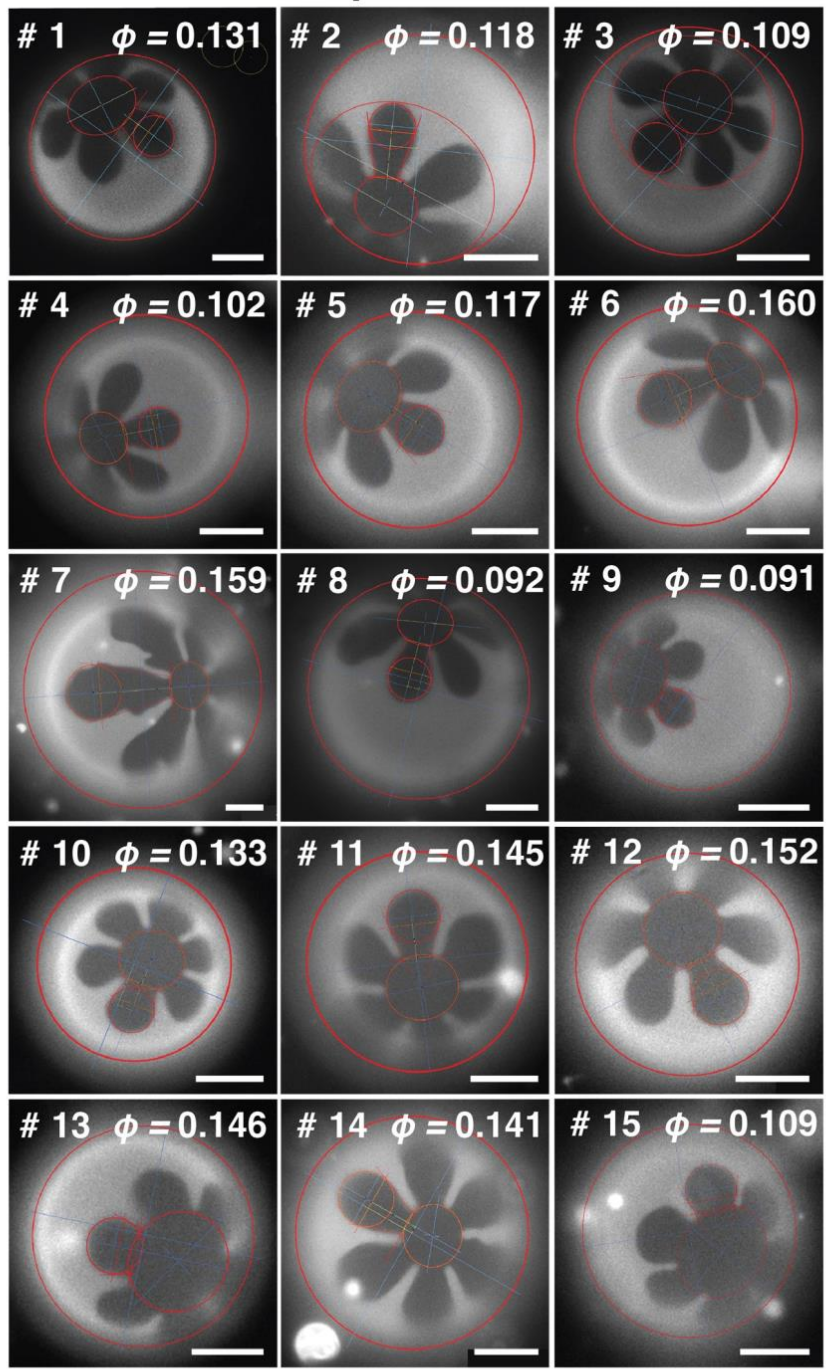

**Figure S8.** Fluorescent images and corresponding solid area fraction for 15 different vesicles with simple flower domains. Embedded lines in each micrograph show the measured quantities based on the previously described methods.

# 3. Supplementary Figures called out in main article

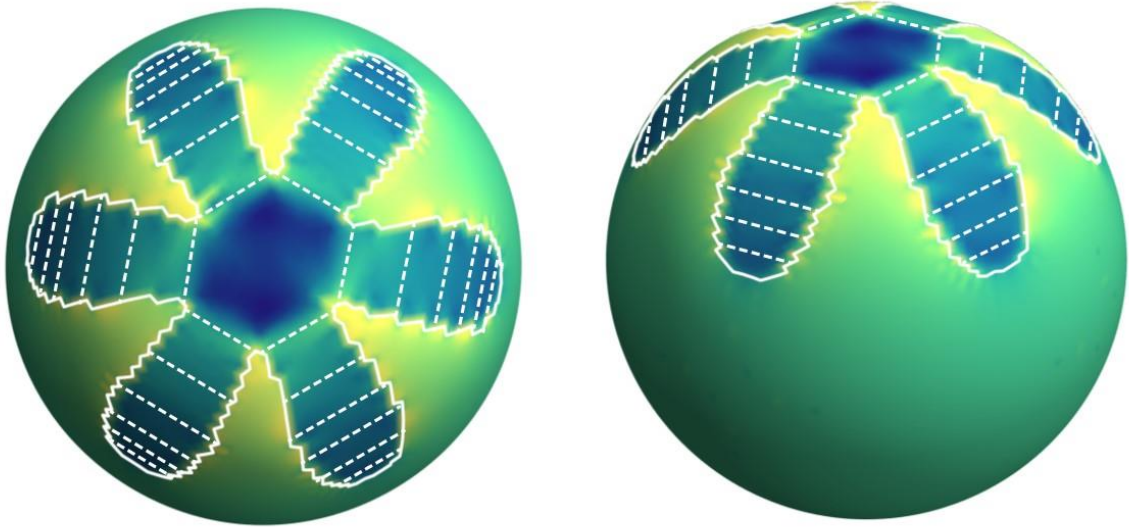

**Figure S9**. Simulated vesicle shape for $\alpha = 3.5$ with mean curvature distribution mapped as in main text Figure 4. Dashed lines illustrate estimates of the generators of the solid domain bending (i.e., the solid surface bends cylindrically around the dashed lines).

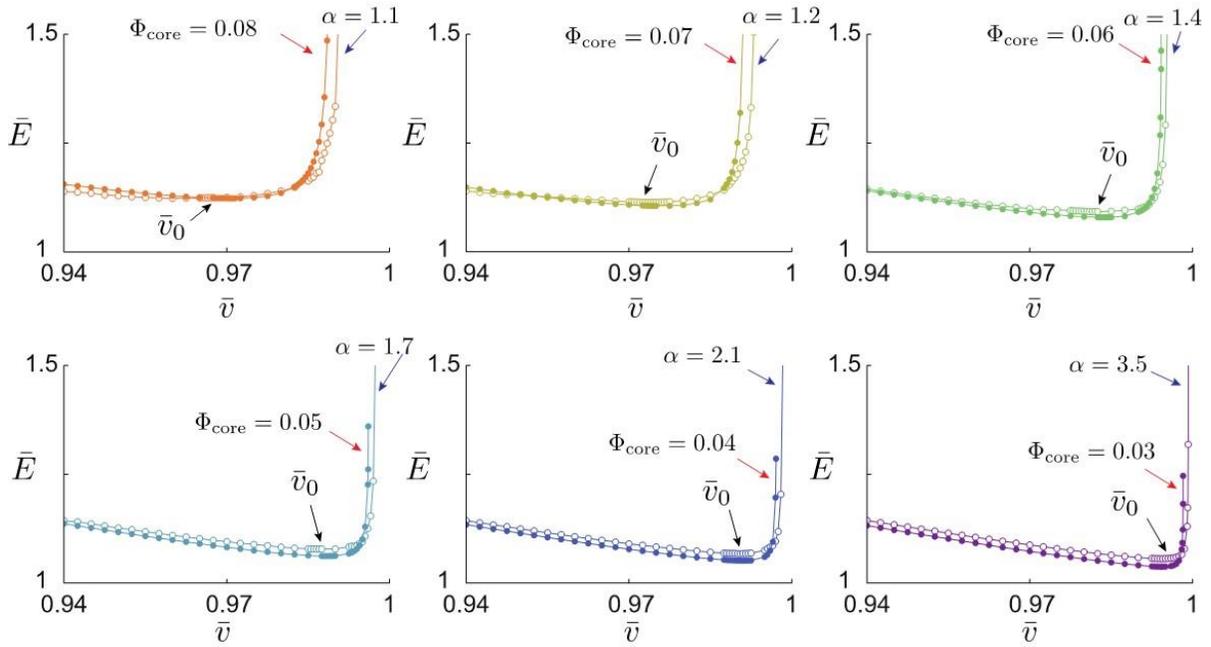

**Figure S10**. Comparison of elastic energy vs. inflation for 14% solid/variable petal size domains to "effective core" models. For each value of $\alpha$, the area fraction of a rigid core $\Phi_{\text{core}}$ is chosen to match the minimal energy value of reduced volume, $\bar{v}_0$. Notably, the elastic energy of flower shaped domains and simpler effective core model match well even far from this vanishing tension point, particularly in the diverging-energy, high-tension regime as vesicles approach maximal inflation.

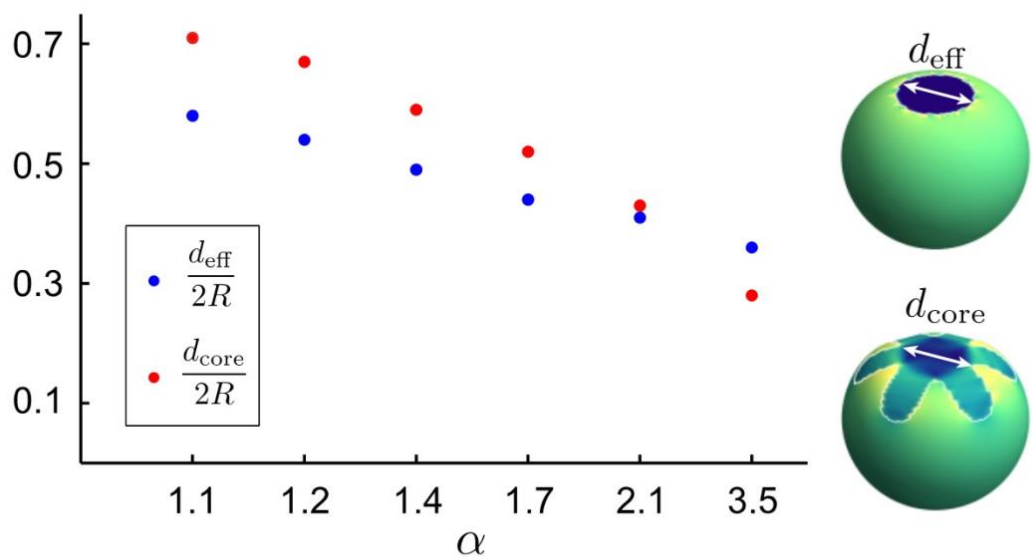

**Figure S11**. Comparison of the size of the effective core region, $d_{\text{eff}}$, to the in-radius size of flowered solid shapes, $d_{\text{core}}$.

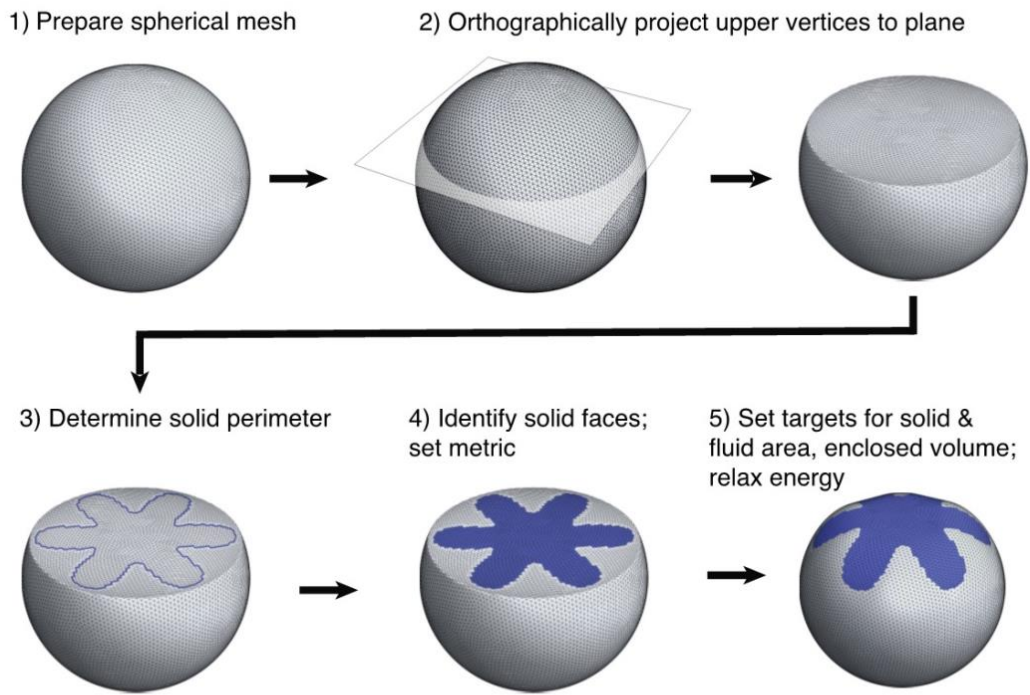

**Figure S12**. Schematic overview of the workflow for Surface Evolver calculations of composite vesicle shapes and energetics.

## 4. Surface evolver model of fluid-solid composite vesicles

Here we describe the procedure to numerically compute optimal elastic energy shapes of composite vesicles. The workflow of our numerical calculation protocol is shown in Figure S12, and summarize the elements of each step as follows:

1) *Initialize spherical mesh* – We initialize a spherical mesh of ~$10^5$ triangular facets.

2) *Flatten upper plane* – To initialize the solid domain, the vertices above a height *z* are orthographically projected down to that plane. The height *z* is chosen to ensure that the solid domain fits on top the planar cut.

3) *Determine the solid perimeter* – A shape for the boundary edge of the solid domain is selected (with a given $\alpha$ value and shape described below) and projected onto the plane at constant *z*, centered around pole of axisymmetry of the "dented sphere".

4) *Prepare solid faces* – Faces interior to the 2D stencil of the domain edge (i.e., with all three vertices interior to the boundary curve) are selected as the solid domain. The data on the facet edge lengths and dot products in this planar state is used to generate the (i.e., flat metric) reference state of the elastic strain energy of the solid (detailed below).

5) *Set targets and relax energy* – Starting from this configuration the elastic energy of the entire vesicle (detailed below) is relaxed in Surface Evolver, targeting constraints on both solid and fluid areas as well as the interval volume.

For the effective core model, the procedure is similar, the exception that circular planar domains are held fixed (i.e., strictly rigid) in the subsequent energy relaxation.

*2D domain shape* - The shape of the solid domain was defined by the following radial function on the planar surface.

$$r(\theta) = r_0 + \frac{1}{2} a r_0 \cos(6\theta) - \frac{1}{10} a r_0 \cos(12\theta) \qquad (4.1)$$

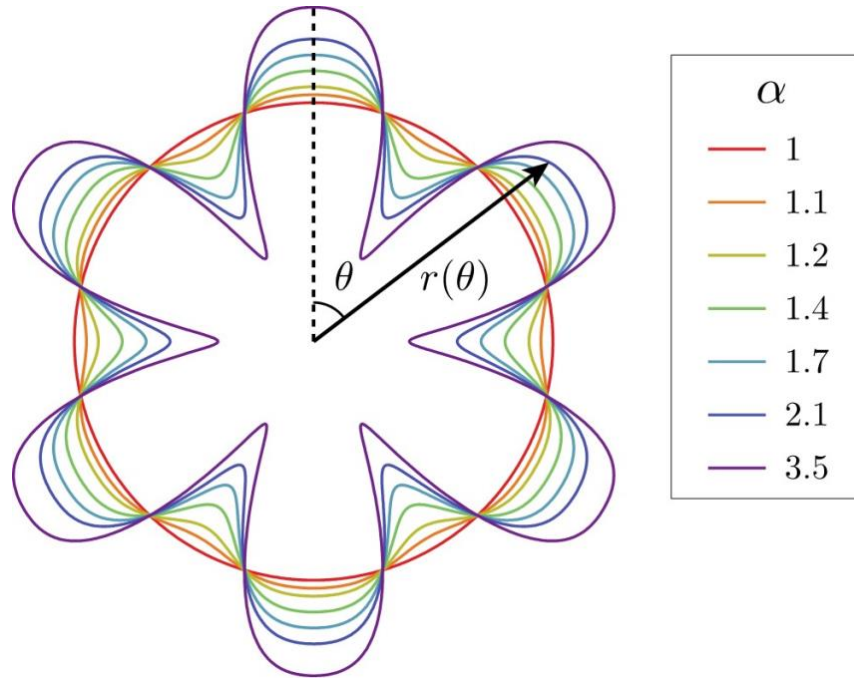

**Figure S13.** Perimeter shapes for 2D solid domains used in Surface Evolver calculations.

which implies $\alpha = r(0)/r(\pi/6) = (5 + 2a)/(5 - 3a)$. The value of $r_0$ was chosen for each $\alpha$ to keep the initial area fraction as close as possible to 14% solid fraction, although subsequent relaxation of the fluid area was needed to more exactly match this ratio. The ratio between the first and second harmonic was selected to give closer approximation to the variably-petaled shapes observed in microscopy over the full range of $\alpha$, shown in Figure S13.

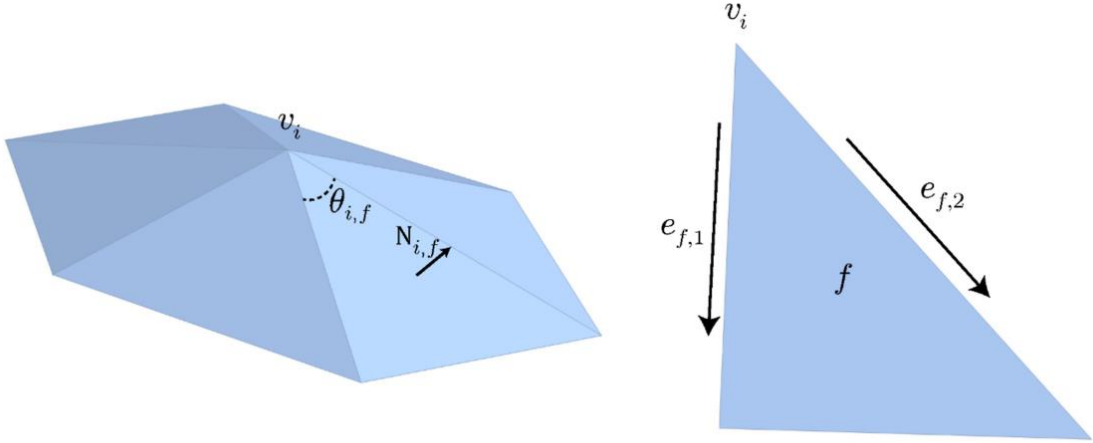

**Figure S14**. Discrete mesh components around a given central vertex $v_i$.

*Mesh geometry* – Based on the mesh geometry defined in Figure S14, we use discrete approximants for mean ($h_i$) and Gaussian ($k_i$) curvature at vertex $\mathbf{v}_i$.[9, 10] The discrete mean curvature follows from the gradient of the area element normal to the mesh,

$$h_i = \frac{1}{2} \frac{\mathbf{F}_i \cdot \mathbf{N}_i}{\mathbf{N}_i \cdot \mathbf{N}_i} \qquad (4.2)$$

where

$$\mathbf{F}_i = \sum_f \boldsymbol{\nabla}_{\mathbf{v}_i} A_{i,f} = \sum_f \frac{1}{4A_{i,f}} \left( -\mathbf{e}_{f,1} |\mathbf{e}_{f,2}|^2 - |\mathbf{e}_{f,1}|^2 \mathbf{e}_{f,2} + (\mathbf{e}_{f,1} \cdot \mathbf{e}_{f,2})(\mathbf{e}_{f,1} + \mathbf{e}_{f,2}) \right) \qquad (4.3)$$

and

$$\mathbf{N}_i = \sum_f \boldsymbol{\nabla}_{\mathbf{v}_i} V_{i,f} = \sum_f \frac{1}{3} A_{i,f} \mathbf{N}_{i,f} \qquad (4.4)$$

is the average normal at the vertex. The Gaussian curvature is simply computed from the sum of the internal angles of the faces at $i$

$$k_i = 2\pi - \sum_f \theta_{i,f} \qquad (4.5)$$

*Elastic energy* - The bending energy at each vertex for both the fluid membrane and the solid domain was computed by the built-in function *star_perp_sq_mean_curvature*,

$$E_b = B_d \sum_i h_i^2 \frac{A}{3} \tag{4.6}$$

where $A_i = \sum_j A_{ij}$ is the total area of adjacent facets and $B_d$ is the discrete bending modulus. Note that in this built-in method, $B_d$ is twice of the continuum bending modulus $B$, i.e., $B_d = 2B$.

For the strain energy at each vertex over the solid domain, we used the built-in function *linear_elastic*,

$$E_s = Y \sum_i \frac{1}{2(1+\nu)} \left( \text{Tr}[C_i^2] + \frac{\nu (\text{Tr}[C_i])^2}{1-\nu} \right) \tag{4.7}$$

$$C_i = \frac{1}{2} (F_i S_i^{-1} - I) \tag{4.8}$$

Where $\nu$ is the Poisson's ratio, $d$ is the dimension, $Y$ is the stretching modulus, and $C_i$ is the Cauchy-Green strain tensor with unstrained Gram matrix $F_i$, strained Gram matrix $S_i$ at each vertex $v_i$, and identity matrix $I$.[11] In our model, $\nu = 0.4$, and the unstrained solid domain configurations are set to be planar (i.e. extracted from the planar configuration in the initialize step 5 in Figure S12 above). The total elastic energy reads

$$E_{\text{elastic}} = E_b^{\text{fluid}} + E_b^{\text{solid}} + E_s^{\text{solid}} \tag{4.9}$$

*Minimization protocol* – Starting from the initial configuration shown in step 5 of Figure S12 above, Surface evolver is used to the minimize the energy while holding the solid and fluid areas at constant ratio and fixing the internal volume to achieve a given target value of $\bar{v}$.[11]

In practice minimization uses a combination of gradient descent (via the command 'g' in Surface Evolver) and Hessian step methods. In Surface evolver, there are two options to apply the Newton's method, *hessian* and *hessian_seek*. *hessian* directly uses the Newton's method and can potentially optimize the energy even faster, but potentially fails (leading to large mesh deformation) if the expansion is not sufficiently good. *hessian_seek* is more stable than *hessian* because it uses the Hessian matrix but it does a search along the gradient direction. In our simulation, *hessian_seek* was mainly used, with occasional trials of *hessian* for particularly sluggish minimizations. To check if an apparent minimum is a saddle point, the command *saddle* is used.

Generally, the minimization proceeds by 100 gradient descent steps followed by a Hessian minimization until the step size falls below $10^{-9}$, and which point the *saddle* command is applied. This procedure is repeated at least three times.

For certain cases the final state fails to reach a minimum (i.e. it is saddle point), or satisfy area and volume constraints. Additionally, if visual inspection of certainly final states shows certain highly distorted mesh regions, or energy differences with nearby parameter values are large, it is suspected that the configuration maybe stuck in a local minimum. For these cases, vertex averaging (for the fluid vertices only) was used to displace vertices and the minimization procedure proceeds. If this fails, the command jiggle is used (for the fluid vertices only) and the minimization procedure proceeds.